\documentstyle[12pt]{article}
\newcommand{\ca}{${\cal A}(2,2)${}}
\newcommand{\cz}{{\cal Z}}
\newcommand{\dmi}{d\mu_{n}(Z)}
\newcommand{\tupn}{\stackrel{(n)}{T}}
\newcommand{\cs}{\zeta}
\newcommand{\csp}{{\zeta}^{\prime}}
\newcommand{\lkwmi}{L^{2}(d\mu_{n})}
\newcommand{\skpr}[2]{\left\langle #1 \, |\, #2 \right\rangle}
\newcommand{\troj}{\triangle_{q_1   q_2}^{j   m}}
\newcommand{\trojpr}{\triangle_{q_{1}^{\prime}   q_{2}^{\prime}}%
^{j^{\prime}   m^{\prime}}}
\newcommand{\del}[2]{\delta_{#1 , #2}  }

\newcommand{\mysection}[1]{\section{\hspace{-20pt}. #1}%
 \setcounter{equation}{0} }
\newcommand{\lew}{\!\!\!\!}
\newcommand{\bm}{\boldmath}
\newcommand{\unbm}{\unboldmath}
\newcommand{\cbm}{\mbox{\bm$C$\unbm}}
 
\begin{document}

\title{QUANTUM $SU(2,2)$-HARMONIC OSCILLATOR}
\author{Wojciech Mulak\em
\\  \\
Institute of Theoretical Physics, University of Wroc{\l}aw \\
Pl. Maxa Borna 9, 50-204 Wroc{\l}aw \\ Poland}
\date{}
\maketitle

\begin{abstract}
The $SU(2,2)$-harmonic oscillator on the phase space ${\cal A}(2,2)=
{SU(2,2)}/{S(U(2)\times U(2))}$ is quantized using the coherent states.
The quantum Hamiltonian is the Toeplitz operator corresponding
to the square of the distance with respect to the $SU(2,2)$-invariant
K\"ahler metric on the phase space. Its spectrum, depending on the choice
of representation of $SU(2,2)$, is computed.
\end{abstract}

\section{Introduction}

The $SU(2,2)$-harmonic oscillator is the generalization of the model
harmonic oscillator with the flat phase space.
In our case the phase space ${\cal A}(2,2)=
{SU(2,2)}/{S(U(2)\times U(2))}\simeq {SO(4,2)}/{SO(4)\times SO(2)}$
is the eight dimensional conformal domain,on which the canonical
coordinates $(x^{\mu},p^{\mu}),\mu =0,...,3$ can be globally
introduced.

The spaces of this type are well known as Cartan
classical domains. They appear in physics and mathematics considered
by many authors. The complex geometry of these spaces and, in
particular, its applications in conformal theories
has been investigated in work of  Coquereaux and Jadczyk
(see [1] and references there). The geometry of
{\ca{}}  is related to the space--time geometry. The Shilov
boundary of {\ca{}} is the compactified
Minkowski space--time, endowed with the
conformal structure of the signature
$(+,-,-,-)$. The compactification
is obtained by addition a light cone at
infinity to the usual Minkowski space--time.

As it is suggested in [2] the conformal domain can be considered
as the replacement of the space-time on the micro scale. This
interpretation is based on the Born's reciprocity idea of the
symmetry between the space-time and the energy-momentum space.
The reciprocity symmetry can be reformulated as the symmetry of
the conformal domain. In the consequence these spaces are not
distinguished on micro scale. The Minkowski space is interpreted as
the very-high-mass,or very-high-energy-momentum-transfer limit of
the conformal domain.

The $SU(2,2)$-harmonic oscillator is the one-body system. It is obtained
from the two-body interacting system by introducing the "center of the
mass" coordinates. The interaction is $SU(2,2)$-invariant. The covariant
harmonic oscillators are used in quark models.In these models the
interaction  between quarks are given by the harmonic oscillator potential.
The model of the relativistic hadron consisting of two quarks interacting
in that way can be
found in [3]. These models have been considered by many authors
(see references in [3]).
It is tempting to interpret our model along the similar
lines.

The quantization by using the Berezin--Weyl calculus [4] provides
the quantum Hamiltonian as the Toeplitz operator [1].
This scheme of quantization involves the system of
Perelomov`s generalized coherent states for $SU(2,2)$
(see [5], [6]). The representation spaces of the
quantization are the Hilbert spaces of the holomorphic
functions on the domain, corresponding to the members of the
discrete series of unitary irreducible representations of
$SU(2,2)$. These representations spaces have
their counterparts in Minkowski space--time
as spaces of distributional boundary values
(see [7]).

By the quantization procedure for different representations
we obtain different spectrum of the quantum Hamiltonian.
In contrast
to the geometric quantization, this quantization does not contain the
prequantization stage. For all representations the quantum Hamiltonian
has discrete and degenerate spectrum. The $SU(1,1)$-harmonic oscillator
has been considered in [8].

We use the $S$--parametrization of {\ca{}} introduced
in [1]. This parametrization provides the description of  the
 geometry of {\ca{}} in terms of its symmetries.

\mysection{SU(2,2) -- HARMONIC OSCILLATOR}

The classical Hamiltonian of the $SU(2,2)$ -- harmonic
oscillator in the S-parametrization
[1] of ${\cal A}(2,2)$
is given by the function:
\begin{equation}\label{1.1}
H=\frac{1}{4} {\rm Tr}({ln}^{2}(S_{0} S))
\end{equation}
This function, up to multiplication constant, is the square
of the distance from the origin $S_0$ of \ca{} , with respect
to the $SU(2,2)$ -- invariant K\"{a}hler metric on \ca{} .
The function (\ref{1.1}) is the generalization of the
Hamiltonian of the harmonic oscillator with the flat phase space.
In the flat case the Hamiltonian of the harmonic
 oscillator can be obtained in this way from the K\"{a}hler metric
 on the phase space  $\Gamma = {\cbm}^N$ .

 \ca{} can be realized as the complex bounded domain
 \begin{equation}\label{1.2}
 {\bf 1} - Z Z^{+}\; > \; 0,
 \end{equation}
 where the points of \ca{} are parametrized by $Z\in M_{2}({\cbm})$.
 Let us introduce the following coordinates on \ca{} given by [7]:
\begin{equation}\label{1.3}
Z=\left| \begin{array}{cc}
          z_{11} & z_{12} \\
	  z_{21} & z_{22}
	  \end{array} \right| = u_1
	  \left| \begin{array}{cc}
          \lambda_{1} & 0 \\
	  0 & \lambda_2
	  \end{array} \right|  u_2
\end{equation}
\begin{displaymath} u_1 = e^{i\phi_{1}\sigma_{1}}
 e^{i\theta_1 \sigma_3} ,\quad
u_2  =  e^{i\theta_2 \sigma_3 }
 e^{i\phi_2 \sigma_1}, \end{displaymath}
\begin{displaymath} \lambda_1= r_{+} e^{i\alpha},\qquad
\lambda_2=r_{-} e^{i\beta}, \end{displaymath}
\medskip
where the Pauli matrices are:
 $\sigma_1 =
 \left(\begin{array}{cc}
1 & 0 \\
0 & -1
\end{array}\right) $ , $\sigma_3 =
 \left(\begin{array}{cc}
0 & -i \\
i & 0
\end{array}\right) $. The conditions for the coordinates are:
\begin{equation}\label{1.4}
0\;\; \leq\; \; r_{+},\: r_{-}\;\; < \;\; 1
\end{equation}
\begin{displaymath} 0\;\;\leq\;\; \theta_1 ,\: \theta_2\;\; \leq\;\;
\frac{\pi}{2} \end{displaymath}
\begin{displaymath} 0\;\; \leq\;\; \alpha ,\;\beta ,\; \phi_1 ,\;
\phi_2 \;\;\leq\;\; 2\pi  \end{displaymath}
Let us express the function (\ref{1.1}) by the variables (\ref{1.4}).
Using the relation between parametrizations we can write:
\begin{equation}
Y \equiv \frac{S_0 S - {\bf 1}}{S_0 S+{\bf 1}}=
\frac{1}{2}
\left|\begin{array}{cc}
Z+Z^{+} & i(Z-Z^{+}) \\
i(Z-Z^{+}) & -(Z+Z^{+})
\end{array}\right|
\end{equation}
Then we have:
\begin{equation}
H=\frac{1}{4}{\rm Tr}\left(ln^2 \left(S_0 S\right) \right)
= \frac{1}{4}{\rm Tr}
\left(ln^2 \frac{{\bf 1}+Y }{{\bf 1} - Y}   \right)
\end{equation}
The Hamiltonian (\ref{1.1}) is invariant under the action
of the isotropy group of the origin $S_0$, then it is invariant
under the transformation
$Z \rightarrow U Z V^{+}=\left|
\begin{array}{cc}
r_{+} & 0 \\
0 & r_{-}
\end{array}\right|  $,
where $U$ and $V$ are the unitary matrices given by the singular value
decomposition for $Z$. Using this invariance we obtain:
\begin{equation}
H\;=\;\frac{1}{2}\left( ln^2 \frac{1+r_{+}}{1-r_{+}} +
ln^2 \frac{1+r_{-}}{1-r_{-}}
\right).
\end{equation}
The canonical variables can be introduced by:
\begin{equation}\label{1.8}
\cz = \left(
x^{\mu} + i\hbar \frac{p^\mu}{p^2}
\right)\sigma_\mu ,
\end{equation}
where the unbounded parametrization of \ca{} by $\cz\in M_{2}({\cbm})$
is used. In this parametrization the condition
(\ref{1.2}) reads:
\begin{equation}\label{1.9}
- i (\cz - \cz^{+})\: > \: 0.
\end{equation}
The relation between (\ref{1.2}) and (\ref{1.9}) parametrizations
is given by
the Cayley transformation:
\begin{equation}
Z\: =\: \frac{{\bf 1} + i\cz }{{\bf 1} - i \cz }
\end{equation}

\mysection{COHERENT STATES FOR SU(2,2)}

Let us consider the discrete series of unitary irreducible
representations of $SU(2,2)$, which are realized in the spaces of
holomorphic functions on \ca{} , namely the representation
of the series $d_0$ in Graev's classification [7].
The members of the series $d_0$ are labeled by the integer
number $n=4,5,\ldots $ and two spin labels $j_1, j_2$. In our
case $j_1=j_2=0$.

Let $|dZ|$ denotes the Euclidean measure on \ca{} . Let $d\mu_n$
denotes the normalized measure given by:
\begin{equation}
\dmi \: = \: N_n\left[ \det ({\bf 1}-Z Z^{+} )
\right]^{n-4}|dZ|\: ,\quad \; n=4,5,\ldots ,
\end{equation}
where $N_n= \frac{(n-3)(n-2)^{2}(n-1)}{\pi^4}$ is the normalization
constant
so that $\int d\mu_n = 1 $. The space of functions on \ca{} :
\begin{equation}
{\cal F}_n = \left\{ f\: {\rm holomorphic}\: :
||f||^{2}_{n}\: = \: \int |f(Z)|^2 \dmi \: < \: \infty \right\}
\end{equation}
is the Hilbert space with the scalar product:
\begin{equation}
\left\langle f | g \right\rangle = \int \overline{f(Z)}g(Z)\:
\dmi\: , \;
f,g \in {\cal F}_n.
\end{equation}
The transformation:
\begin{equation}\label{2.4}
\left(\tupn (g)f\right)(Z) \: = \:
\left[\det ( C Z + D)\right]^{-n}  f((A Z + B)(C Z + D)^{-1})\: ,
\end{equation}
\begin{displaymath}
f\in {\cal F}_n \: , \;\quad g^{-1}=\left|
\begin{array}{cc}
A & B \\
C &  D
\end{array}\right|\in SU(2,2)\: , \end{displaymath}
defines the unitary irreducible representation of $SU(2,2)$ in
${\cal F}_n$. The system of coherent states of type
$( \tupn ,|\Psi_0 >\: =\: 1 )$ is obtained by an action of representation
(\ref{2.4}):
\begin{equation}\label{2.5}
\tupn | \Psi_0 \:> = \left[ \det ( C Z + D )\right]^{-n}\: ,
\;\quad 1 = |\Psi_0 >\: \in {\cal F}_n .
\end{equation}
The states obtained by (\ref{2.5}) can be parametrized by points of \ca{} :
\begin{equation}
|\cs >\: =\: \frac{\left[\det({\bf 1}-\cs^{+}\cs) \right]^{n/2} }%
{\left[\det({\bf 1}-\cs^{+} Z) \right]^{n} }\:,\;\;\quad
{\bf 1 } -\cs\cs^{+}\: >\: 0.
\end{equation}
The family $\left\{ |\cs >\: :{\bf 1 } -\cs\cs^{+}\: >\: 0  \right\}$
forms the system of generalized coherent states for $SU(2,2)$
[6]. It has the property of the resolution of the unity:
\begin{equation}
N_{n} \int |\cs >\; < \cs | d\mu (\cs ) \:=\: {\bf 1}_{{\cal F}_{n}} \: ,
\end{equation}
where
$ d\mu (\cs )= [\det ({\bf 1} - \cs\cs^{+})]^{-4}|d\cs |$
is the $SU(2,2)$-invariant measure on \ca{} . Every state
$|\Psi >\: \in {\cal F}_n$ has the continuous representation:
\begin{equation}\label{2.8}
|\Psi >\: \longrightarrow \: < \cs |\Psi > \: =\: C_{\Psi}(\cs )\: =\:
\left[\det({\bf 1}-\cs^{+}\cs) \right]^{n/2} \Psi ( \cs ).
\end{equation}
The representation (\ref{2.8}) has the property:
\begin{equation}
C_{\Psi}(\cs^{\prime})=\int K(\cs^{\prime},\cs )C_{\Psi}(\cs )
d\mu (\cs )\: ,
\end{equation}
where
\begin{equation}
K(\csp ,\cs )\: =\: N_n <\csp | \cs >\: =
N_n  \frac{
\left[\det ({\bf 1} - \cs^{\prime +}\csp )\right]^{n/2}
\left[\det ({\bf 1} - \cs^{+}\cs )\right]^{n/2}
  }{\left[\det ({\bf 1} - \cs^{+}\csp )\right]^{n} }
\end{equation}
is the reproducing kernel:
\begin{equation}
K(\cs , \cs^{\prime\prime})\: =\:\int K(\cs ,\csp )K(\csp ,
\cs^{\prime\prime})\: d\mu (\csp ).
\end{equation}

\mysection{QUANTUM SU(2,2)-HARMONIC OS\-CI\-LLATOR}

The classical system on \ca{} can be quantized
by using the Berezin-Weyl calculus.
This scheme of the quantization involves the system of coherent
states.

For every representation $\tupn$ of $SU(2,2)$, $n=4,5,\ldots $
we obtain different quantizations in the representation spaces
${\cal F}_n$. The operator corresponding to the classical
 observable is its Toeplitz operator constructed by using the
 generalized Bergman projection.

 Let $\lkwmi$ denotes the Hilbert space of the measurable
 and square integrable functions on \ca{} with respect to the measure
  $d\mu_{n}$. The generalized Bergman projection [4]:
 \begin{equation}
 P_{B} : \lkwmi\longrightarrow {\cal F}_n\: ,
 P_{B}^{+}=P_B =P_{B}^{2}
 \end{equation}
 is given by:
 \begin{equation}
 (P_B\: f)(Z) \: =\: \int L_{n}(Z,\cs )f(\cs ,\cs^{+})d\mu_{n}(\cs )\: ,
 \quad f\in\lkwmi \: ,
 \end{equation}
 where $L_{n}(\csp ,\cs )= [\det ( {\bf 1}-\cs^{+}\csp )]^{-n}$
 is the generalized Bergman kernel.
 The quantization associates to each function
 $f\in \lkwmi$ an operator $\hat{f}$ in ${\cal F}_n$ [1]:
\begin{equation}\label{3.3}
f\longrightarrow \hat{f}=N_{n}\int f(\cs ,\cs^{+})\: |\:\cs >\:
<\cs |\:d\mu (\cs ).
\end{equation}
Acting by $\hat{f}$ on $|\Psi >\in {\cal F}_n$ we have:
\begin{equation}
\hat{f}\, |\, \Psi\, >\: =\: P_{B}(f\cdot\Psi ).
\end{equation}
Then the operator (\ref{3.3}):
\begin{equation}
\hat{f}=P_{B}\circ f\circ P_B
\end{equation}
is the Toeplitz operator corresponding to the function f.

Let us describe the orthonormal base in ${\cal F}_n$.
The base consists of the functions [7]:
\begin{equation}\label{3.6}
\triangle^{jm}_{q_1   q_2}(Z)\: =\:
({\cal N}^{j   m})^{-1} (\det Z)^{m}D^{j}_{q_1   q_2}(Z) \:  ,
\end{equation}
\begin{displaymath}
\; m\: =\: 0,1,2,\ldots \quad 2 j\: =\: 0,1,2,\ldots\quad
-j\:\leq\: q_1\: ,  q_2\:\leq\: j
\end{displaymath}
where the function $D^{j}_{q_1   q_2}$ is the extension of the polynomial
well known from the $SU(2)$ representation theory:

\begin{equation}
D^{j}_{q_1   q_2}(Z)\: =\:
\left[\frac{(j+q_1 )! (j-q_1 )!}{(j+q_2)! (j-q_2 )!}\right]^{1/2}\quad
\sum_{s=\max (0,q_1 +q_2 )}^{s=\min (j-q_2 , j+q_2 )}
{j+q_2 \choose s}\times
\end{equation}
\begin{displaymath}
{}\times{j-q_2 \choose s-q_1-q_2}
z_{11}^{s} z_{12}^{j+q_1 - s} z_{21}^{j+q_2 -s} z_{22}^{s-q_1 - q_2}
\end{displaymath}
and the normalization constant is given by:
\begin{equation}
\left( {\cal N}^{jm}  \right)^2 =
(n-1)(n-2)^{2}(n-3)
\frac{(n-3)! (n-4)! (m+2j +1)! m!}
{(2j+1)(m+n-2)!(m+2j+n-1)!}.
\end{equation}
The orthonormality of (\ref{3.6}) reads:
\begin{equation}
\skpr{\troj}{\trojpr}\: =\:
\del{j^{\prime}}{j}\del{m^{\prime}}{m}\del{q_{1}^{\prime}}{q_1}
\del{q_{2}^{\prime}}{q_2}
\end{equation}
By the quantization (\ref{3.3}) of the $SU(2,2)$-harmonic oscillator
we obtain
the quantum Hamiltonian:
\begin{equation}\label{3.10}
\hat{H}\: =\: P_{B}\circ H\circ P_B\: ,\quad H\in \lkwmi
\end{equation}
In order to find the spectrum of the operator (\ref{3.10})
let us compute the matrix element:
\begin{equation}
\skpr{\trojpr}{\hat{H}\troj}\: =\:
\skpr{\trojpr}{H\troj}
\end{equation}
In this order we use the coordinates (\ref{1.3}). After some
calculations
we obtain:
\begin{equation}
\skpr{\trojpr}{\hat{H}\troj}\: =\:
\del{j^{\prime}}{j}\del{m^{\prime}}{m}
\del{q_{1}^{\prime}}{q_{1}}
\del{q_{2}^{\prime}}{q_2}
\stackrel{\lew (n)}{E^{jm}_{q_1   q_2}}
\end{equation}
\begin{displaymath} \stackrel{\lew (n)}{E^{jm}_{q_1   q_2}}\: =\:
\skpr{\troj}{\hat{H}\troj} \end{displaymath}
Then the operator $\hat{H}$ is diagonal in the base
(\ref{3.6}), while
its eigenvalues are $\stackrel{\lew (n)}{E^{jm}_{q_1   q_2}}$:
\begin{equation}
\hat{H}\troj \: = \: \stackrel{\lew (n)}{E^{jm}_{q_1   q_2}} \troj
\end{equation}
The eigenvalues are given by the integral:
\begin{equation}\label{3.14}
 \stackrel{\lew (n)}{E^{jm}_{q_1   q_2}} \; = \;
 \alpha_{j,m,n}\sum_{q=-j}^{q=j}
 \sum_{i=0}^{n-4}\sum_{l=0}^{n-4}
 (-1)^{i+l}{n-4 \choose i}{n-4 \choose l}\times
\end{equation}
\begin{displaymath}
{} \times \int_{0}^{1} dr_{+}\int_{0}^{1} dr_{-}
 \left( ln^2 \frac{1+r_{+}}{1-r_{+}}
 + ln^2 \frac{1+r_{-}}{1-r_{-}}
  \right) r_{+}^{2(j+m+q+i)+1} r_{-}^{2(j+m-q+l)+1}
  \left( r_{+}^2 - r_{-}^2 \right)^2 ,
\end{displaymath}
where
\begin{displaymath}
    \alpha_{j,m,n} =\frac{(m+n-2)! (m+2j+n-1)!}{
    (2j+1)(n-3)!(n-4)!m!(m+2j+1)!}
\end{displaymath}

Let us denote for $N=0,1,2,\ldots $
\begin{equation}
S_{1}(N)\equiv \frac{1}{N+1}\sum_{a=0}^{N} \frac{1}{2 a +1}
\end{equation}
\begin{displaymath} S_{2}(N) \equiv \left\{ \begin{array}{ccl}
               \frac{1}{N+1}\sum_{b=1}^{N} \sum_{a=b}^{N}
	       \frac{1}{2 a +1}\cdot\frac{1}{2b} & , & N=1,2,\ldots \\
               {}&{}& \\
	       0 & , & N=0 \\
	       \end{array} \right.
\end{displaymath}
\medskip
\begin{displaymath}
 S(N)\equiv S_{1}(N) ln 2 + S_{2}(N)
\end{displaymath}
The integral (\ref{3.14}) can be computed using the formula:
\begin{equation}
\int_{0}^{1} ln^2 \frac{1+r}{1-r} \ \cdot \ r^{2N+1} dr = 4 S(N)
\end{equation}
The eigenvalues are given by the formula:\pagebreak[4]
\begin{equation}
\stackrel{\lew (n)}{E^{jm}_{q_1   q_2}}  =
\frac{4}{(2j+1)(n-3)! }\sum_{i=0}^{n-4} (-1)^i {n-4 \choose i}\times
\end{equation}
\begin{displaymath}
{} \times  \left\{ \frac{(m+2j+n-1)!}{(m+2j+1)! }
\left[ (m+n-2)S(2j +m+2+i) \right. \right.
\end{displaymath}
\begin{displaymath}
\left.  -(m+1)S(2j+m+1+i)\right] +
\end{displaymath}
\begin{displaymath}
\left. {}  +  \frac{(m+n-2)!}{m!} \left[ (2j+m+2)S(m+i) -
(2j+m+n-1)S(m+1+i)\right] \right\}
\end{displaymath}
We observe that the eigenvalue does not depend on $q_1$, $q_2$
indices. Then the eigenvalue $\stackrel{\lew (n)}{E^{jm}}$
is $(2j+1)^2$ degenerate.

\mysection{Remarks}

The result of the quantization depends on the choice of
representation of $SU(2,2)$. The question arises how to interpret
this choice. According to the Berezin's interpretation [4]
the number of representation depends on parameter $h$,
which plays the role of the Planck constant. By taking the
limit $h\rightarrow 0$ the correspondence
principle is obtained. From this point of view the relation
between this parameter and the Planck constant in (\ref{1.8})
is not clear.

The Hamiltonian of the $SU(2,2)$-harmonic oscillator
may also be interpreted as the generalization of the Born's quantum
metric operator, which plays the crucial role in the reciprocity
theory.
This fact may encouraged us to interpret the spectrum
of the quantum Hamiltonian in the spirit of this
theory.

\bigskip

\underline{Acknowledgments}

The author is grateful to Prof. A. Jadczyk
for inspiration, suggestions and helpful discussions,
as well as would like to thank Prof. R. Coquereaux
for critical comments.

\end{document}